\documentclass[prb,twocolumn,floatfix]{revtex4-1}
\usepackage{latexsym}
\usepackage{graphicx}
\usepackage{subfigure}

\usepackage{subfigure}
\usepackage{array}
\usepackage{verbatim}
\usepackage{amsmath}
\usepackage{color}
\usepackage{xcolor}
\usepackage{soul}
\usepackage{tabu}
\usepackage{multirow}
\usepackage{lipsum}
\usepackage[normalem]{ulem}
\begin{document}
\graphicspath{{fig/}{./}}
\title{Transition to an excitonic insulator from a two-dimensional conventional insulator}
\author{Efstratios Manousakis$^{(1,2)}$}
\affiliation{$^{(1)}$ Department  of  Physics,
  Florida  State  University,  Tallahassee,  Florida  32306-4350,  USA\\
$^{(2)}$Department   of    Physics,   University    of   Athens,
  Panepistimioupolis, Zografos, 157 84 Athens, Greece}
\date{\today}
%\section{Introduction}
\begin{abstract}
  In this work, first, we present a general formulation to investigate
  the ground-state and elementary excitations of an excitonic insulator
  (EI) in real materials.   In addition, we discuss the out-of-equilibrium state induced (albeit transiently) by high-intensity light illumination of a conventional two-dimensional (2D) insulator.
  We then, present various band-structure models which allow us to
  study the transition
  from a conventional insulator to an EI in 2D
  materials as a function of the dielectric constant, the conventional insulator gap (and chemical potential), the bandwidths of the conduction and valence bands
  and the Bravais lattice unit-cell size.
  One of the goals of this investigation
  is to determine which range of these experimentally determined parameters
  to consider in order to find the best candidate materials to realize the
  excitonic insulator.
  The numerical solution to the EI gap equation for various band-structures
  shows a significant and interesting momentum-dependence of the
  EI gap function $\Delta(\vec k)$ and of the zero-temperature
  electron and hole momentum-distribution  across the Brillouin zone.
  Last, we discuss the fact that these features can be detected by tunneling microscopy.

\end{abstract}
\maketitle
\section{Introduction}

The instability of the conventional insulating state to that
of the so-called excitonic-insulator (EI) in  semiconductors was pointed out
and studied a long-time ago\cite{NFMott,Knox,PhysRev.158.462,RevModPhys.40.755,RevModPhys.42.1}. Within the Bose-Einstein condensation (BEC)
framework, the EI state can occur
when the electron–hole binding energy exceeds the charge bandgap; when this
happens, clearly the conventional-insulator ground-state becomes unstable.
The ground-state exciton population is then determined by balancing the
negative exciton formation energy against the pair-exciton
repulsion energy.
It has been argued that in bulk materials the EI state
can occur in small-bandgap semiconductors and
small bandwidth semimetals\cite{RevModPhys.40.755}.
This phenomenon has also been investigated in connection with
various aspects of the 
quantum many-boson problem\cite{PhysRevLett.74.1633,PhysRevLett.126.027601,Eisenstein2004,science.aam6432,Butov_2004,Baranov2012}; the presence of
repulsion among exciton pairs can  stabilize   
superfluid  and crystalline  phases  by means of suppression of  density  and
phase fluctuations\cite{Baranov2012,LOZOVIK2007399,PhysRevLett.110.145302}.
Spectroscopic evidence for EIs in various materials
has  been   reported\cite{science.aam6432,PhysRevLett.99.146403,PhysRevB.90.155116,Du2017,pnas.2010110118}, however,  conclusive  evidence   for  the
existence of such a state of matter remains elusive.
In recent experimental work in transition metal dichalcogenide (TMD)
double layers, it was argued\cite{Ma2021} that a
quasi-equilibrium of spatially indirect exciton fluid can be established
when the
bias voltage applied between the  two electrically isolated TMD layers
was tuned to a range that  populates bound electron–hole pairs, but not
free electrons  or holes\cite{PhysRevLett.121.067702,PhysRevB.102.085154,PhysRevLett.120.177702}.

While the  EI was originally conceived more than
half a century ago, during the last decade several different families of
two-dimensional (2D) or quasi-2D insulating or semiconducting materials
have been synthesized;  it might be reasonable to expect that
once  the relevant parameters of a properly chosen material are fine-tuned,
such a material can  become the host of this unique
state of matter. The reason for holding this expectation for these pure 2D or quasi-2D materials is two fold. First, there are already reports\cite{science.aam6432,PhysRevLett.99.146403,PhysRevB.90.155116,Du2017,pnas.2010110118,Wang2019} that show spectroscopic evidence
for the existence of the EI state. Second, 
one expects a smaller average dielectric constant in single
and double atomic layers due to the presence of vacuum or of a physical
gap surrounding these layers. Examples of such materials are, carbon-based
and related exfoliated 2D materials, the TMD mentioned
above and organic–inorganic hybrid perovskites\cite{Stoumpos2016,doi:10.1126/science.aac7660} where quantum wells can
be engineered with chemical growth techniques\cite{Gao2019}.
The latter 2D
perovskite structures can be understood as atomically thin slabs
that are cut from the three-dimensional parent structures along
different crystallographic directions, which are sandwiched by two layers of
large organic cations\cite{Saparov2016}.

In the present work, we describe a general formulation based on the
linearization of the electron-hole interaction and by subsequent diagonalization
of the resulting Bogoliubov-de Gennes Hamiltonian matrix. Then, we derive an
analytic expression for a BCS-like gap-equation of the EI state.
We  iteratively solve the
resulting gap equation to determine the full momentum dependence of the gap function, the EI ground-state electron and hole momentum distribution and the
quasiparticle excitations of the excitonic state.
We use a variety of interesting band-structure cases of two-dimensional
materials and provide full numerical solution.  Last, we study the
transition to the EI state from the conventional-insulator
state as a function of the parameters, dielectric constant, band-structure
details including the bandwidth, the effective conventional-insulator
gap and the Bravais lattice unit-cell size. This study is instructive
for the search for the appropriate 2D materials which might be 
best suited to host this unconventional insulating state.
In addition, we find some band-structures which host a more exotic
EI state with regard to the bare-electron momentum distribution
in the EI ground-state.  This distribution displays peaks at non-trivial points inside the Brillouin zone that can play the role of a smoking gun for
confirming the presence of the EI state by means of tunneling
experiments\cite{doi:10.1073/pnas.2207449119,PhysRevB.101.085142,PhysRevLett.120.177702}.

The paper is organized as follows. In Sec.~\ref{sec:formulation}
we present the formulation of the problem and its analytical aspects
and we derive the gap equation for the most general case.
In 
Sec.~\ref{sec:features} we discuss the general features of the
problem and how to find the best approach to search for materials which
host the EI state. In 
Sec.~\ref{sec:solution} we provide numerical solutions to the gap
equation by iteratively solving the EI gap equation.
In Sec.~\ref{EItransition} we study the transition from the conventional-insulator to the EI state as a function of the band-structure parameters,
the dielectric constant, etc. Last, in 
Sec.~\ref{sec:Discussion} we give a summary of our results and conclusions
and we discuss how to experimentally detect the EI state with its various features.

\iffalse
  In a Dirac or Weyl semimetal the RPA dielectric function diverges in the
  $q \to 0$ limit, and, thus, these systems cannot become excitonic
  insulators. When there is a gap $G$ opening up in these topological
  materials, there is a chance for
  such a state to occur. However, in the case of a small gap, the dielectric pol-arizability would be proportional to $1/G$ and, so, if $G$ is small the dielectric function would be large
  and the electron-hole interaction would be weak. On the other hand,
\fi
  
\section{Formulation}
\label{sec:formulation}

Let us start from the interaction term between electrons excited in the
conduction band $\beta$ and those in the valence band $\alpha$ in the simpler
case where the interaction does not cause interband transitions, i.e., 
\begin{eqnarray}
 && \hat V_{\mathrm{eff}} = {1 \over {2}}\sum_{\alpha\beta\vec k_1,\vec k_2,\vec q}
  V^{\alpha\beta}_{\vec k_1\vec k_2}(\vec q)     c^{\dagger}_{\vec k_1+\vec q,\alpha} c^{\dagger}_{\vec k_2-\vec q,\beta}  c_{\vec k_2,\beta} c_{\vec k_1,\alpha}, \\
 && V^{\alpha\beta}_{\vec k_1\vec k_2}(\vec q) \equiv \langle \vec k_1 \alpha; \vec k_2 \beta | \hat V | \vec k_1+\vec q \alpha; \vec k_2 - \vec q \beta\rangle,
    \label{inter-exciton}
\end{eqnarray}
where $| \vec k \alpha\rangle $ is the Bloch state normalized in the
volume $\Omega$ of the crystal and $\hat V$ is the Coulomb interaction
screened by the dielectric matrix.
%We can define as a reference state, the ground-state $|\Psi_0 \rangle$ of
%the conventional insulator.
Let us also define the hole creation and annihilation operators as
\begin{eqnarray}
  h^{\dagger}_{-\vec k,\alpha} \equiv c_{\vec k,\alpha}, \hskip 0.2 in 
  h_{-\vec k,\alpha} \equiv c^{\dagger}_{\vec k,\alpha},
\end{eqnarray}
in terms of which and after changing  dummy summation momentum label,
the interaction term above takes the following form:
\begin{eqnarray}
  \hat V_{\mathrm{eff}} = -{1 \over {2} } \sum_{\alpha\beta\vec k_1,\vec k_2,\vec q}
  V^{\alpha\beta}_{\vec k_1\vec k_2}(\vec q)    h^{\dagger}_{\vec k_1+\vec q,\alpha}  c^{\dagger}_{\vec k_2-\vec q,\beta}  c_{\vec k_2,\beta} h_{\vec k_1,\alpha} .
  \label{inter-exciton4}
\end{eqnarray}
This form can be obtained from Eq.~\ref{inter-exciton} by replacing the
creation/annihilation operator of valence electrons by the hole creation/annihilation operator. Notice that the electron-hole interaction has the
opposite sign, which means that there is an attractive interaction between
electrons of the conduction band and holes in the valence band. This attractive
interaction leads to formation of bound-states of electron-hole pairs which
behave as bosons and which can form a BEC. In the weak coupling limit, this
phenomenon can be approached as a BCS pairing state. To see how the latter
can be described, we use the Bogoliubov-Valatin factorization approach for the interaction quartic term   as follows:
\begin{eqnarray}
 && h^{\dagger}_{\vec k_1+\vec q,\alpha}  c^{\dagger}_{\vec k_2-\vec q,\beta}  c_{\vec k_2,\beta} h_{\vec k_1,\alpha} \to \langle h^{\dagger}_{\vec k_1+\vec q,\alpha}  c^{\dagger}_{\vec k_2-\vec q,\beta} \rangle  c_{\vec k_2,\beta} h_{\vec k_1,\alpha} \nonumber \\
  &+&
  h^{\dagger}_{\vec k_1+\vec q,\alpha}  c^{\dagger}_{\vec k_2-\vec q,\beta} \langle  c_{\vec k_2,\beta} h_{\vec k_1,\alpha} \rangle.
  \end{eqnarray}
Using momentum conservation (i.e., ignoring Umklapp processes) we
obtain:
\begin{eqnarray}
  \hat V_{\mathrm{eff}} &\to&  -  \sum_{\vec k\alpha\beta}
  \Bigl ( \Delta_{\alpha\beta} (\vec k)  c_{\vec k \beta} h_{-\vec k\alpha}
  + h.c. \Bigr ), \\
  \Delta_{\alpha,\beta}(\vec k)  &\equiv&
  {1 \over {2} }\sum_{\vec q}   V^{\alpha\beta}_{\vec k,-\vec k}(\vec q) \eta^*_{\alpha\beta} (\vec k + \vec q),
  \label{inter-exciton-linearized}\\
  \eta^*_{\alpha\beta}(\vec k) &\equiv& \langle h^{\dagger}_{\vec k,\alpha}  c^{\dagger}_{-\vec k,\beta} \rangle.
  \label{eq:eta}
\end{eqnarray}
The treatment of the resulting Hamiltonian
\begin{eqnarray}
  \hat H_{\mathrm{eff}} &=& - \sum_{\vec k\alpha} \epsilon_{\alpha}(\vec k) h^{\dagger}_{\vec k \alpha }
    h_{\vec k \alpha} +
    \sum_{\vec k\beta} \epsilon_{\beta}(\vec k) c^{\dagger}_{\vec k \beta }
    c_{\vec k \beta}\nonumber \\
    &-& \sum_{\vec k\alpha\beta}
  \Bigl ( \Delta_{\alpha\beta} (\vec k)  c_{\vec k \beta} h_{-\vec k\alpha}
  + h.c. \Bigr ),
\end{eqnarray}
where $\epsilon_{\alpha}(\vec k) = E_{\alpha}-\mu_{\alpha}$ (and
$\epsilon_{\beta}(\vec k) = E_{\beta}-\mu_{\beta}$
is standard. Notice that we have allowed for a different chemical potential
between the valence ($\mu_{\alpha}$) and conduction ($\mu_{\beta}$) bands. This difference may take into
account the case where there is photoexcitation of the sample
by means of laser-light illumination. The
chemical potential difference $\delta \mu = \mu_{\beta} - \mu_{\alpha}$
controls the coexistence of electrons in the conduction band and holes
in the valence band. This will occur for a transient time.
In the case where the self-consistently determined value of $\Delta_{\alpha\beta}$ is non-zero, there would be pairing of the electron-hole system.

Using the Bogoliubov-de Gennes      matrix
\begin{eqnarray}
  {\bf M}_{\alpha\beta}(\vec k) \equiv 
  \begin{pmatrix}
    -\epsilon_{\alpha}(\vec k) & 0 & 0 &  -\Delta^*_{\alpha\beta}(\vec k) \\
    0 & \epsilon_{\beta}(\vec k) &  \Delta^*_{\alpha\beta}(\vec k) & 0 \\
    0 &  \Delta_{\alpha\beta}(\vec k) & \epsilon_{\alpha}(\vec k) & 0  \\
      -\Delta_{\alpha\beta}(\vec k) & 0 & 0 & -\epsilon_{\beta}(\vec k) 
  \end{pmatrix},
\end{eqnarray}
the above Hamiltonian can be cast in the following form
\begin{eqnarray}
 &&\hat H_{\mathrm{eff}} =  \sum_{\vec k} {\cal H}_{\mathrm{eff}}(\vec k) + C_0, \\
    &&   {\cal H}_{\mathrm{eff}}(\vec k) = {1 \over 2 } {\vec c}^{\dagger}_{\vec k}
       {\bf M}_{\alpha\beta}(\vec k) {\bf c}_{\vec k}\\
   &&    {\bf c}^{\dagger}_{\vec k} =  (
           h^{\dagger}_{-\vec k \alpha} 
             c^{\dagger}_{\vec k \beta} 
             h_{-\vec k \alpha}
             c_{\vec k \beta}), \\ 
        &&  C_0 = { 1 \over 2} (-\epsilon_{\alpha}(\vec k)
       + \epsilon_{\beta}(\vec k) + \Delta_{\alpha\beta}(\vec k)
       + \Delta^*_{\alpha\beta}(\vec k)).
\end{eqnarray}
The process of diagonalization of this Hamiltonian
is described in the Appendix~\ref{AppendixA}. After carrying out the
diagonalization, $\hat H_{\mathrm{eff}}$ takes the diagonal form
  \begin{eqnarray}
    {\hat H}_{\mathrm{eff}} = \sum_{\vec k} \Bigl [ {\cal E}_1(\vec k)
    {\cal C}^{\dagger}_{\vec k} {\cal C}_{\vec k} +
    {\cal E}_2(\vec k)
    {\cal H}^{\dagger}_{\vec k} {\cal H}_{\vec k} \Bigr ] + E_0,
  \end{eqnarray}
  where
\begin{eqnarray}
  {\cal E}_{1}(\vec k) &=& \bar{\epsilon}_{\alpha\beta} 
  + {\cal R}_{\alpha\beta}
  \hskip 0.2 in {\cal E}_{2}(\vec k) = \bar{\epsilon}_{\alpha\beta}  - {\cal R}_{\alpha\beta}, \\
   {\cal R}_{\alpha\beta} &\equiv& 
   \sqrt{( \delta \epsilon_{\alpha\beta} )^2 + |\Delta_{\alpha\beta}(\vec k)|^2},
   \label{calr}\\
  \bar{\epsilon}_{\alpha\beta} &\equiv& {{\epsilon_{\alpha}(\vec k) + \epsilon_{\beta}(\vec k)} \over 2}, \hskip 0.01 in
  \delta \epsilon_{\alpha\beta} \equiv {{\epsilon_{\alpha}(\vec k) - \epsilon_{\beta}(\vec k)} \over 2}, 
\end{eqnarray}
and
\begin{eqnarray}
  {\cal C}^{\dagger}_{\vec k} &=& \kappa_{-} c^{\dagger}_{\vec k \beta} - \chi_- h_{-\vec k\alpha}, \label{eq:quasia} \\
  {\cal H}_{-\vec k} &=& \kappa_{+} c^{\dagger}_{\vec k \beta} - \chi_+  h_{-\vec k\alpha}, \label{eq:quasib} \\
  \kappa_{\pm} &=& {{\delta\epsilon_{\alpha\beta} \pm {\cal R}_{\alpha\beta}}
    \over {{\cal D}^{\pm}_{\alpha\beta}}}, \hskip 0.2 in 
    \chi_{\pm} = {{\Delta_{\alpha\beta}}     \over {{\cal D}^{\pm}_{\alpha\beta}}}.\\
       {\cal D}^{\pm}_{\alpha\beta} &\equiv& \sqrt{(\delta\epsilon_{\alpha\beta}\pm{\cal R}_{\alpha\beta})^2+|\Delta_{\alpha\beta}|^2},
\end{eqnarray}
where ${\cal C}^{\dagger}_{\vec k}$ corresponds to the ${\cal E}_1$ eigenvalue and
  when $\Delta \to 0$, the operator ${\cal C}^{\dagger}_{\vec k} \to c^{\dagger}_{\vec k \beta}$.
  The ${\cal H}_{-\vec k}$ operator corresponds to the ${\cal E}_2$
    eigenvalue and
  when $\Delta \to 0$, ${\cal H}_{-\vec k} \to h_{-\vec k \beta}$.
  The other two solutions which corresponds to the $-{\cal E}_1$ and $-{\cal E}_2$ eigenvalues are ${\cal C}_{\vec k}$ and ${\cal H}^{\dagger}_{-\vec k}$ operators
  which are the adjoint operators. 

  Now the interacting ground-state $| \Psi_0 \rangle$ is defined as follows:
\begin{eqnarray}
  {\cal C}_{\vec k } | \Psi_0 \rangle &=& 0, \hskip 0.2 in
  {\cal H}_{\vec k } | \Psi_0 \rangle = 0.
  \label{eq:vacuum}
\end{eqnarray}
In order to calculate $\eta_{\alpha\beta}(\vec k)$ given by Eq.~\ref{eq:eta} above
as an expectation value of 
\begin{eqnarray}
  \eta_{\alpha\beta}(\vec k) = \langle \Psi_0|
  h^{\dagger}_{\vec k,\alpha}  c^{\dagger}_{-\vec k,\beta} | \Psi_0 \rangle,
\end{eqnarray}
with respect to the interacting ground-state, we 
invert Eqs.~\ref{eq:quasia},\ref{eq:quasib} (see Appendix~\ref{AppendixA}).
Using Eqs.~\ref{eq:vacuum} it is straightforward to show that
\begin{eqnarray}
  \eta_{\alpha\beta}(\vec k) = {{\Delta_{\alpha\beta}(\vec k)} \over {2
      {\cal R}_{\alpha\beta}}}.
\end{eqnarray}
Therefore, the gap equation takes the form:
\begin{eqnarray}
  \Delta_{\alpha,\beta}(\vec k)  =
        {1 \over {4 } } \sum_{\vec k'}   V_{\alpha\beta}(|\vec k -\vec k'|)
       {{\Delta_{\alpha\beta}(\vec k')} \over {
           {\cal R}_{\alpha\beta}(\vec k')}},
           \label{eq:gap}
\end{eqnarray}
where ${\cal R}_{\alpha\beta}$ is defined by Eq.~\ref{calr}. Notice that we have assumed that the effective interaction depends only on the momentum transfer.
This equation needs to be solved iteratively to determine the gap function
$\Delta_{\alpha\beta}(\vec k)$ from the functions $\epsilon_{\alpha}(\vec k)$,
$\epsilon_{\beta}(\vec k)$, $V_{\alpha\beta}(|\vec k-\vec k'|)$, which are
assumed to be given.
%We will use just one valence and one conduction band

As we will show, it is interesting to calculate the ground-state
electron momentum distribution in the EI state.
The electron momentum distribution, i.e., the ground-state
expectation value of the bare electron and hole number operator is given as
\begin{eqnarray}
  n_{\alpha}(\vec k) &=& \langle \Psi_0|
  h^{\dagger}_{\vec k,\alpha}  h_{\vec k,\alpha} | \Psi_0 \rangle =
  {{[(\delta\epsilon_{\alpha\beta}+ {\cal R}_{\alpha\beta})D^{(-)}_{\alpha\beta}]^2} \over {4 ({\cal R}_{\alpha\beta}\Delta_{\alpha\beta})^2}}, \nonumber \\
  n_{\beta}(\vec k) &=& \langle \Psi_0|
  c^{\dagger}_{\vec k,\beta}  c_{\vec k,\beta} | \Psi_0 \rangle =
  {{[D^{(+)}_{\alpha\beta}]^2} \over {4 {\cal R}^2_{\alpha\beta}}}.
  \label{eq:md}
\end{eqnarray}
These are the coherence factors that can be accessed by tunneling experiments
and this is described in Sec.~\ref{sec:Discussion}.

\begin{figure*}
    \vskip 0.3 in \begin{center}
        \subfigure[]{
            \includegraphics[scale=0.5]{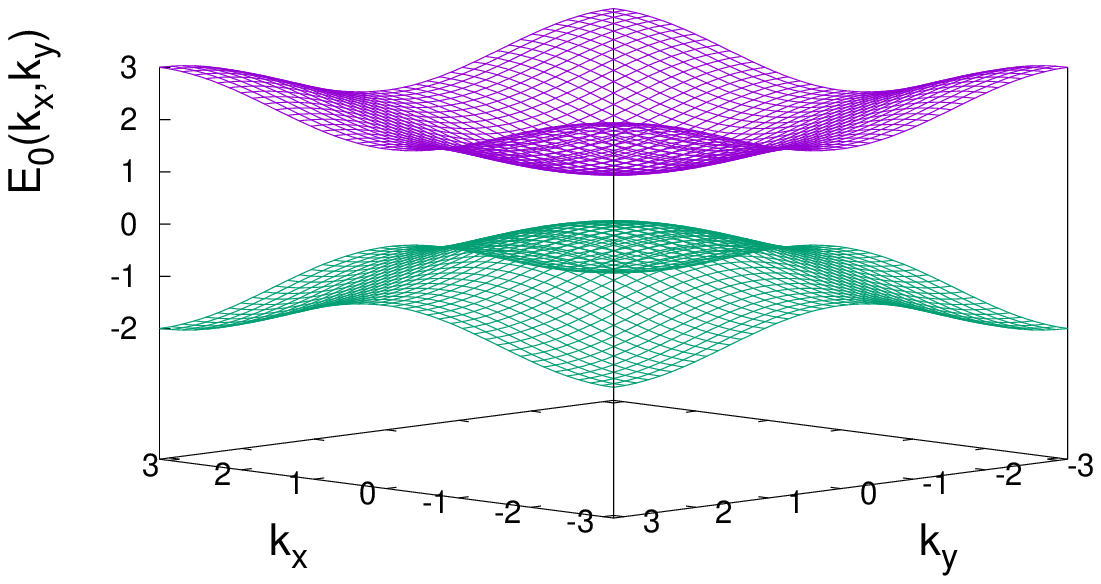} 
           \label{fig1A}
        }
        \subfigure[]{
            \includegraphics[scale=0.5]{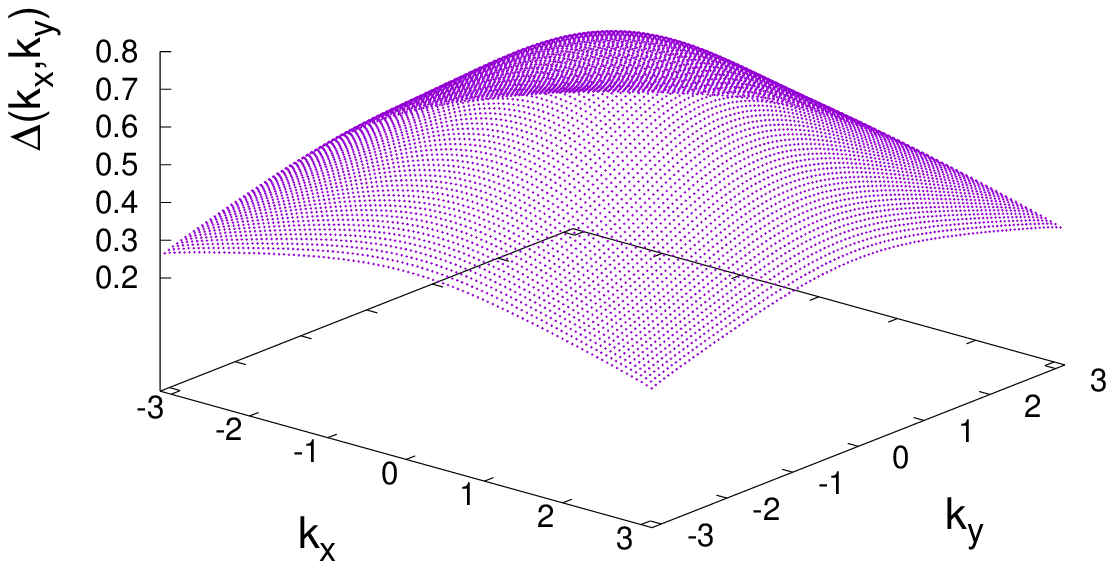}
            \label{fig1B}
        }
        \\
        \subfigure[]{
            \includegraphics[scale=0.5]{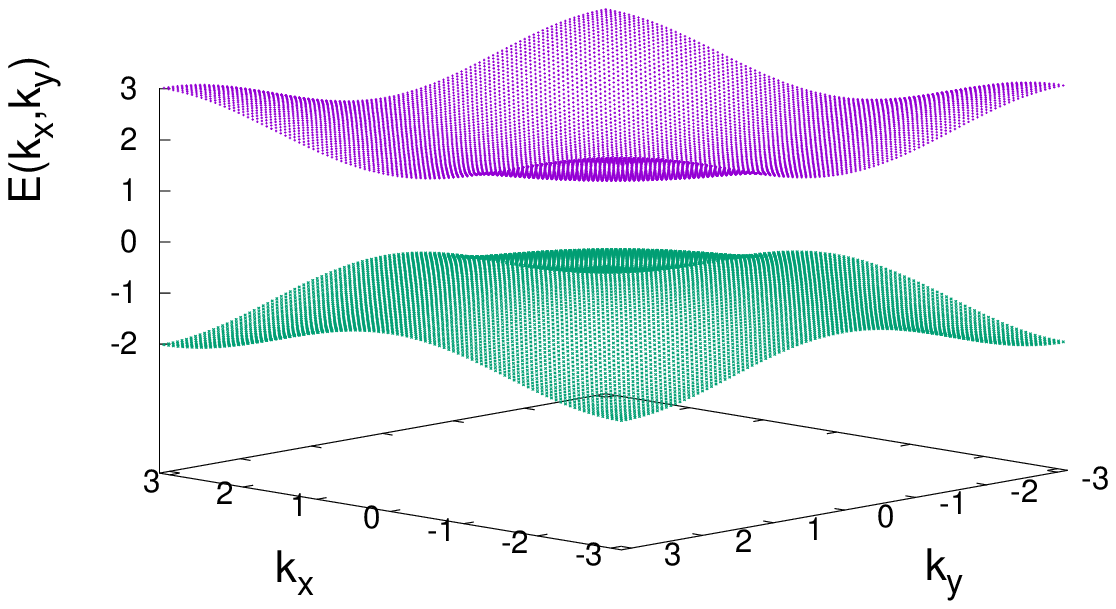}
            \label{fig1C}
        }       
        \subfigure[]{
            \includegraphics[scale=0.5]{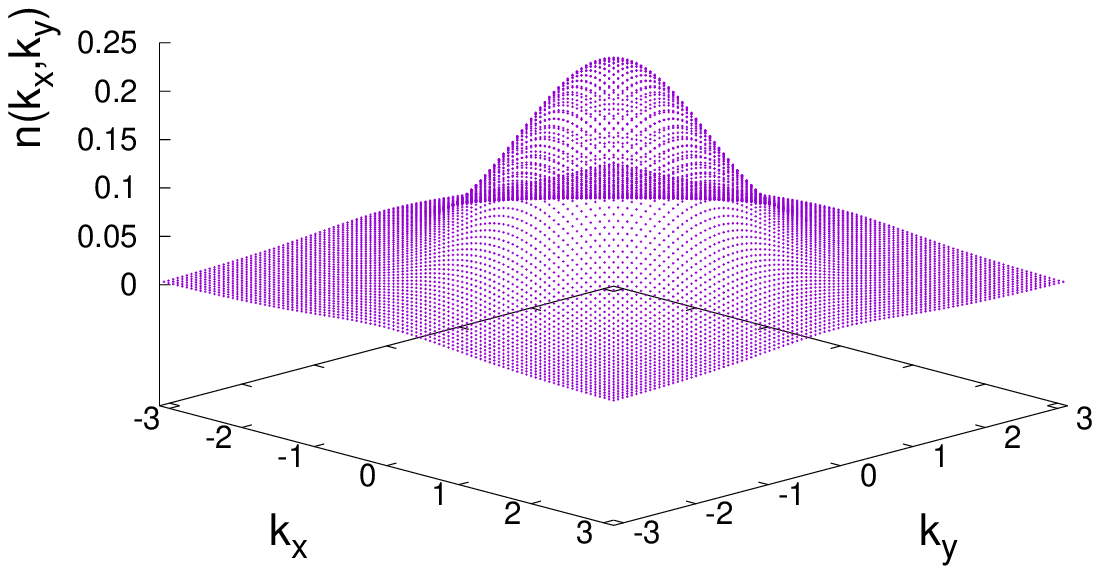}
           \label{fig1D}
        }
    \end{center}
\caption{(a) The non-interacting valence and conduction bands (relative to their corresponding chemical potentials) in the case of the
      band-structure given by Eq.~\ref{simple.direct} for $\gamma=1.0$ eV  and bandwidth $W=$2 eV. (b) The pairing gap  and  (c) the quasiparticle energy (relative to the chemical potentials)  are shown for the case (a) using $\epsilon=3$ and $a=5 a_0$. d) The ground-state momentum distribution of electrons in the conduction-band  for this case.}
\vskip 0.2 in
\end{figure*}

\begin{figure*}
    \vskip 0.3 in \begin{center}
        \subfigure[]{
            \includegraphics[scale=0.5]{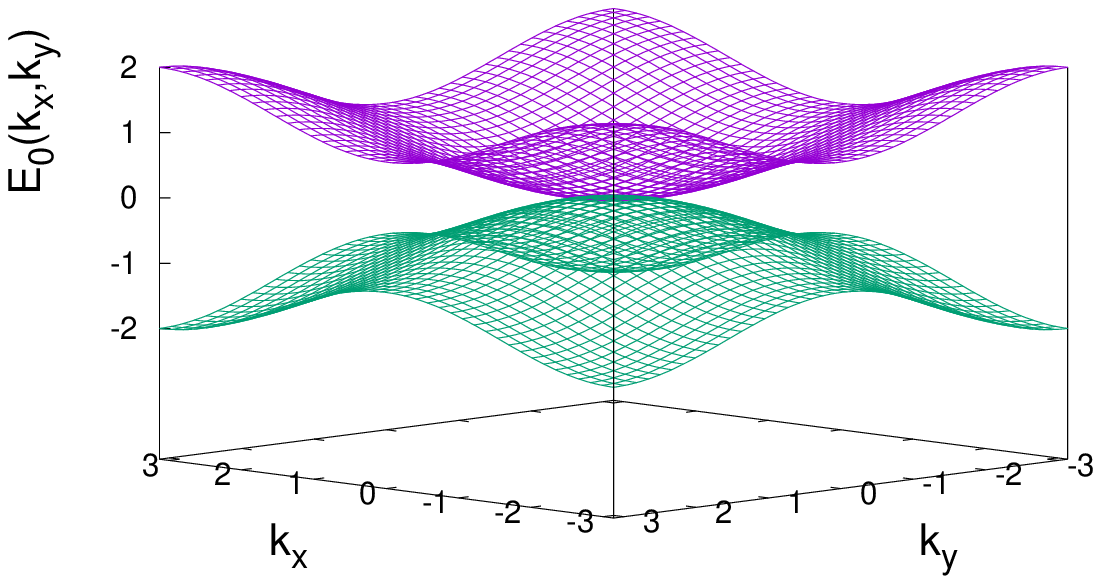} 
           \label{fig2A}
        }
        \subfigure[]{
            \includegraphics[scale=0.5]{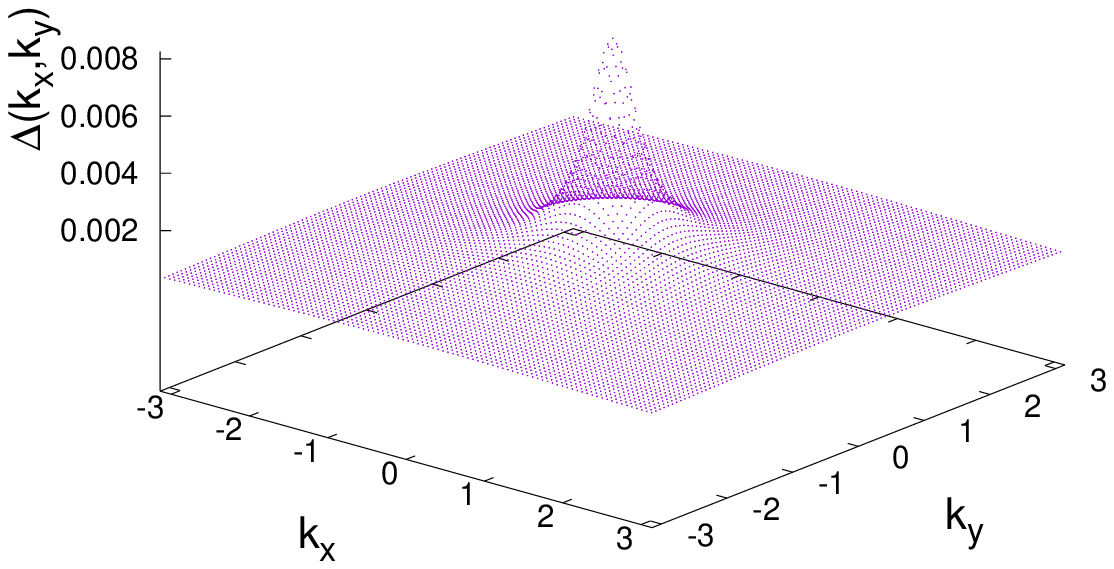}
            \label{fig2B}
        }\\
        \subfigure[]{
            \includegraphics[scale=0.5]{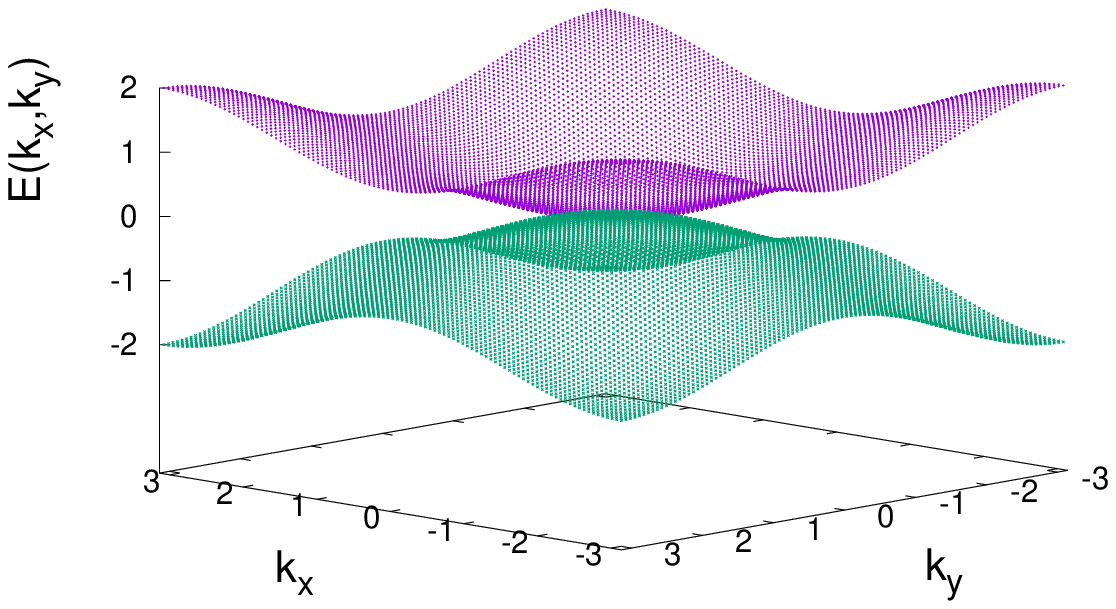}
           \label{fig2C}
        }       
        \subfigure[]{
            \includegraphics[scale=0.5]{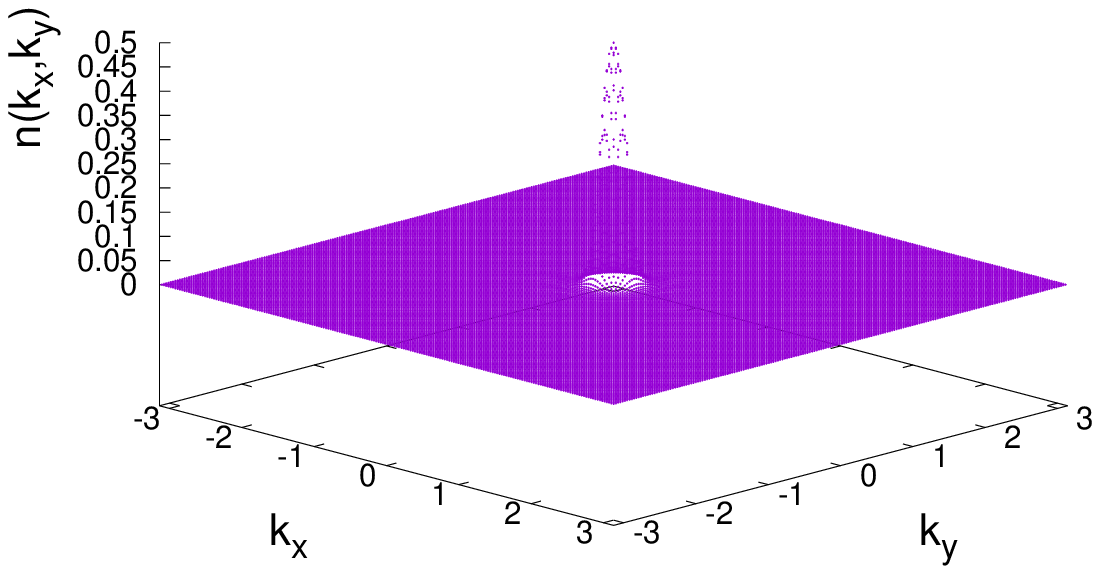}
           \label{fig2D}
        }       
    \end{center}
    \caption{(a) The non-interacting valence and conduction bands (relative to their corresponding chemical potentials) in the case of the
      band-structure given by Eq.~\ref{simple.direct} is shown for $\gamma=0$ eV  and bandwidth $W=$2 eV. (b), (c) and (d) illustrate the  calculated pairing gap, the quasiparticle energy (relative to the chemical potentials) and the ground-state momentum distribution of electrons in the conduction-band respectively, using $\epsilon=10$ and $a=20 a_0$.}
      \vskip 0.2 in
\end{figure*}

\begin{figure*}
    \vskip 0.3 in \begin{center}
        \subfigure[]{
            \includegraphics[scale=0.5]{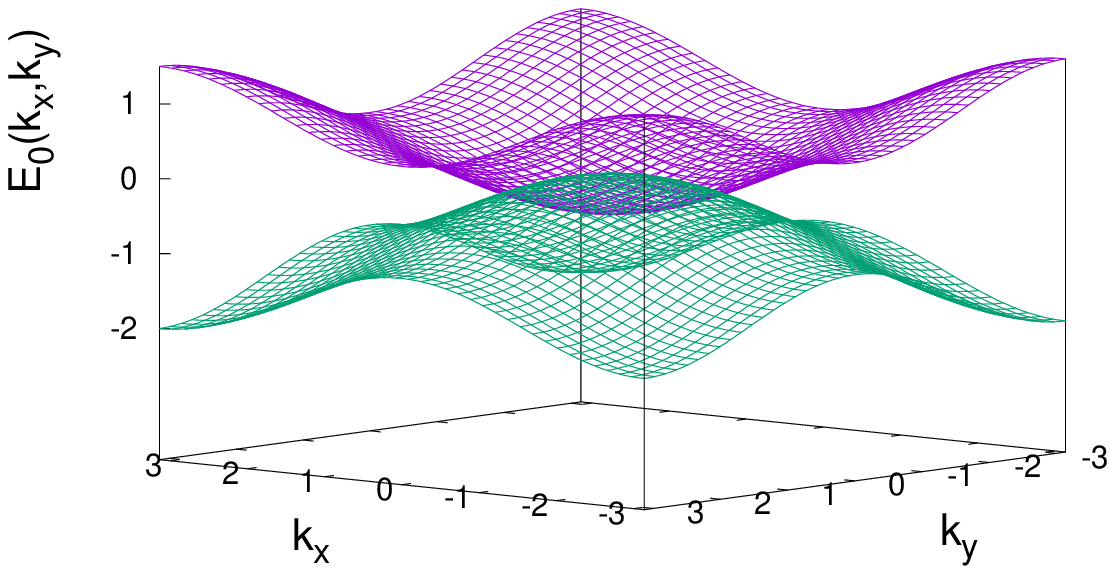} 
           \label{fig3A}
        }
        \subfigure[]{
            \includegraphics[scale=0.5]{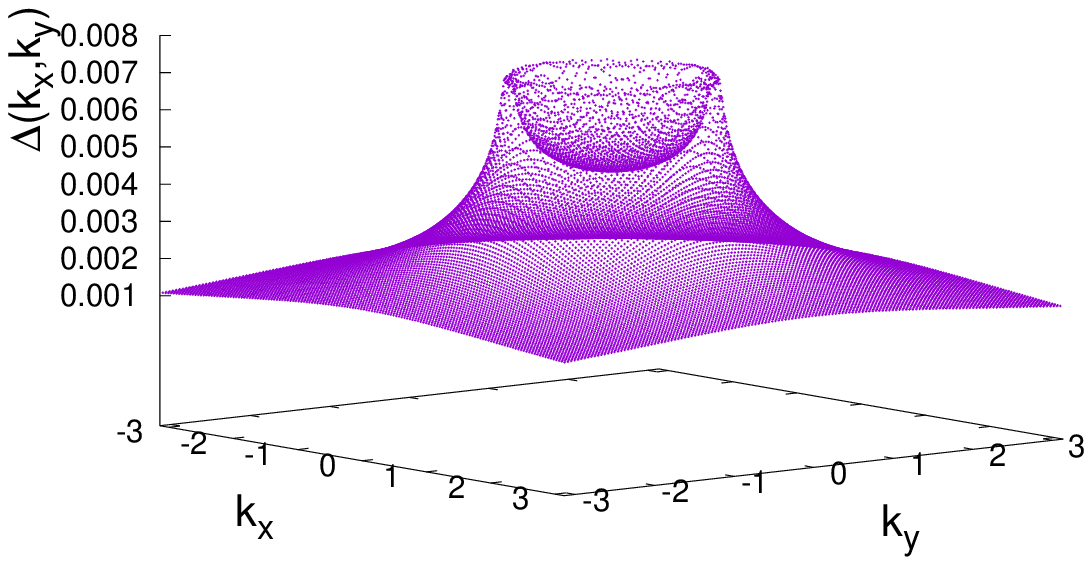}
            \label{fig3B}
        }\\
        \subfigure[]{
            \includegraphics[scale=0.5]{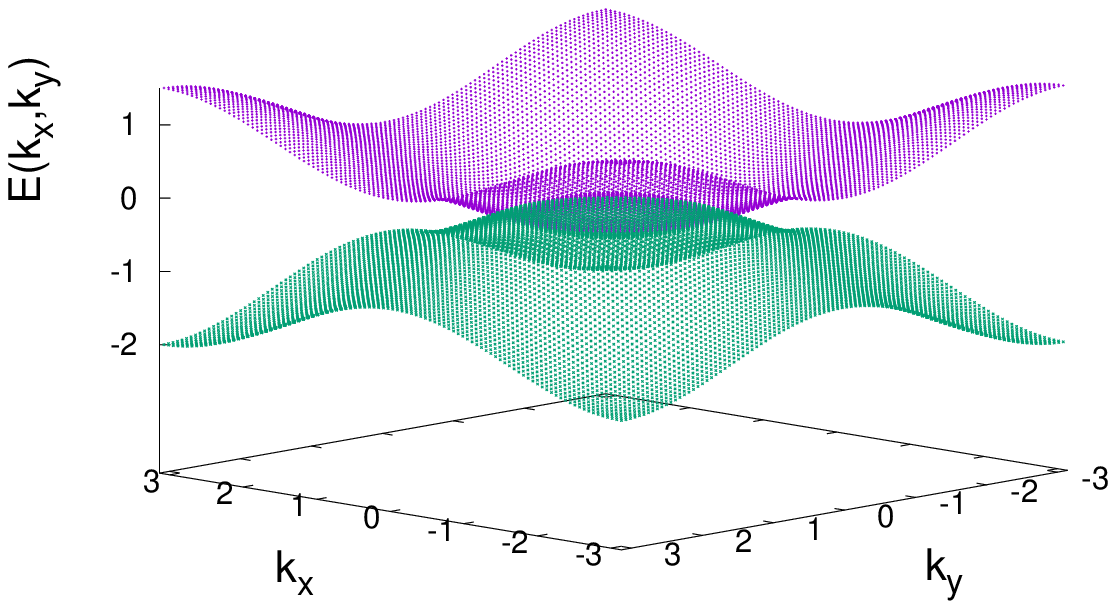}
           \label{fig3C}
        }       
        \subfigure[]{
            \includegraphics[scale=0.5]{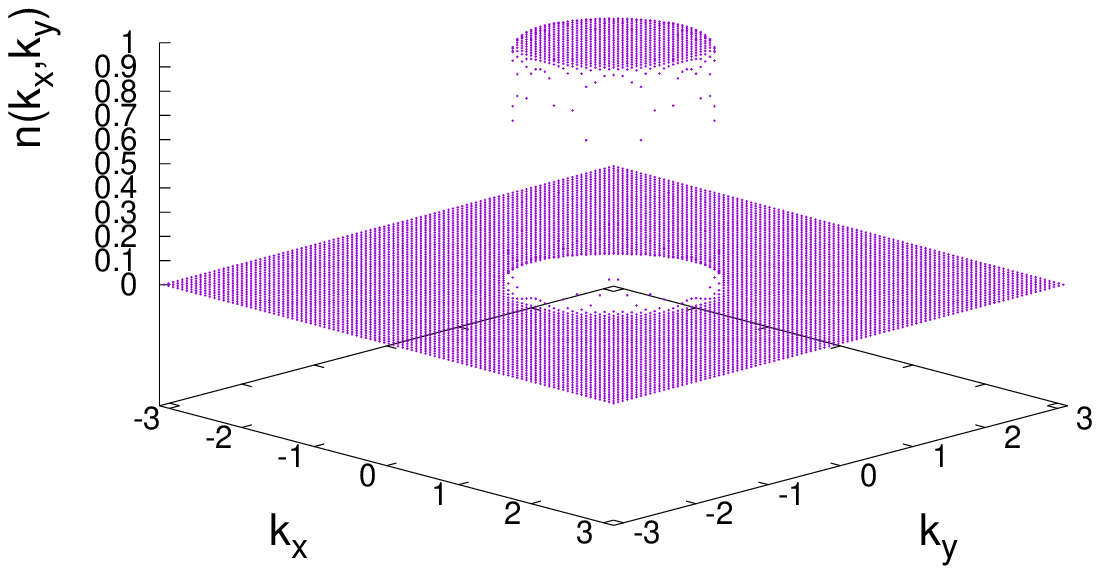}
           \label{fig3D}
        }
    \end{center}
    \caption{(a) The non-interacting valence and conduction bands (relative to their corresponding chemical potentials) in the case of the
      band-structure given by Eq.~\ref{simple.direct} is shown for $\gamma=-0.5$ eV  and bandwidth $W=$2 eV. (b), (c) and (d) illustrate the  calculated pairing gap, the quasiparticle energy (relative to the chemical potentials) and the ground-state momentum distribution of electrons in the conduction-band respectively, using $\epsilon=10$ and $a=20 a_0$.}
\vskip 0.2 in
\end{figure*}

\begin{figure*}
    \vskip 0.3 in \begin{center}
        \subfigure[]{
            \includegraphics[scale=0.6]{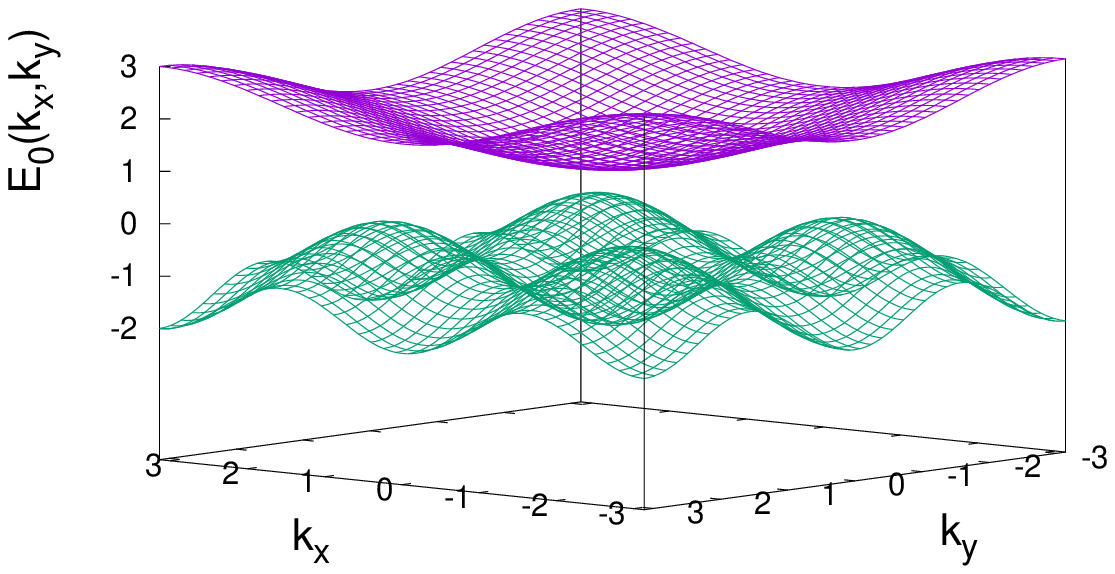} 
           \label{fig4A}
        }
        \subfigure[]{
            \includegraphics[scale=0.6]{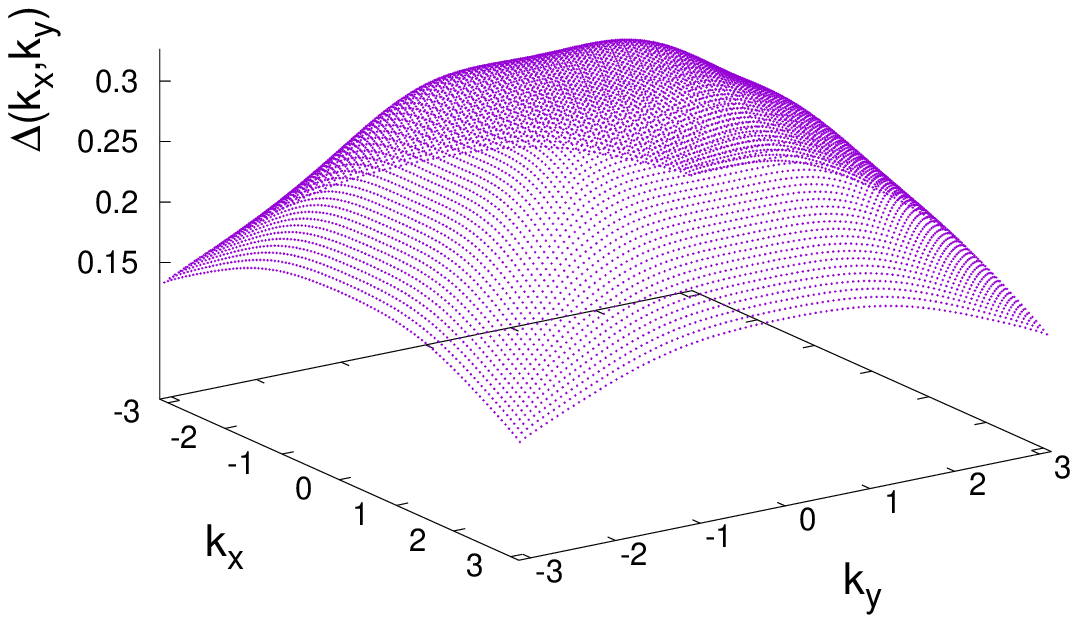}
            \label{fig4B}
        } \\
        \subfigure[]{
          \includegraphics[scale=0.6]{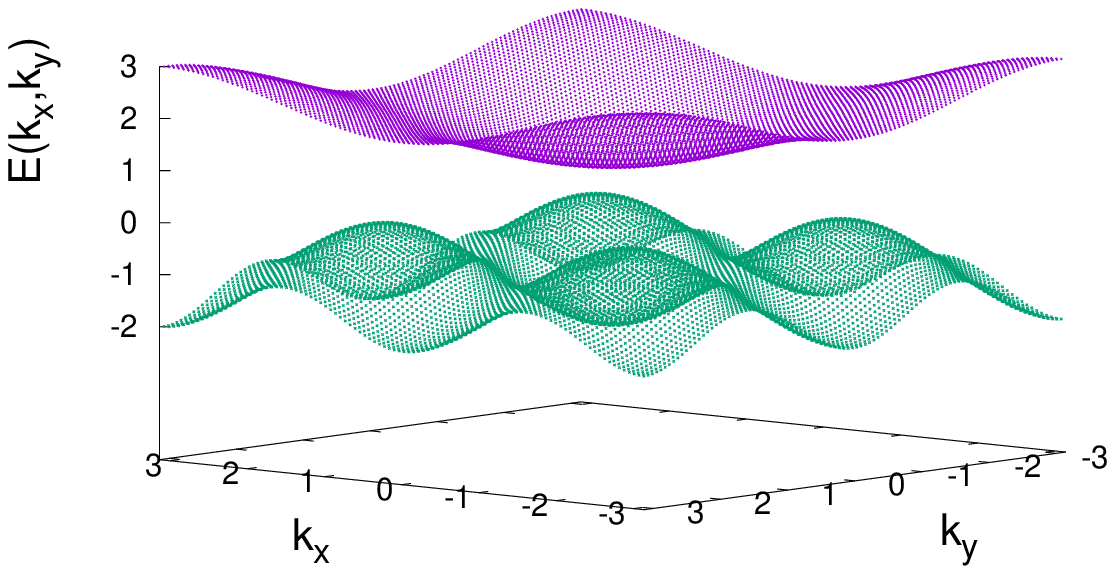}
           \label{fig4C}
        }
        \subfigure[]{
            \includegraphics[scale=0.6]{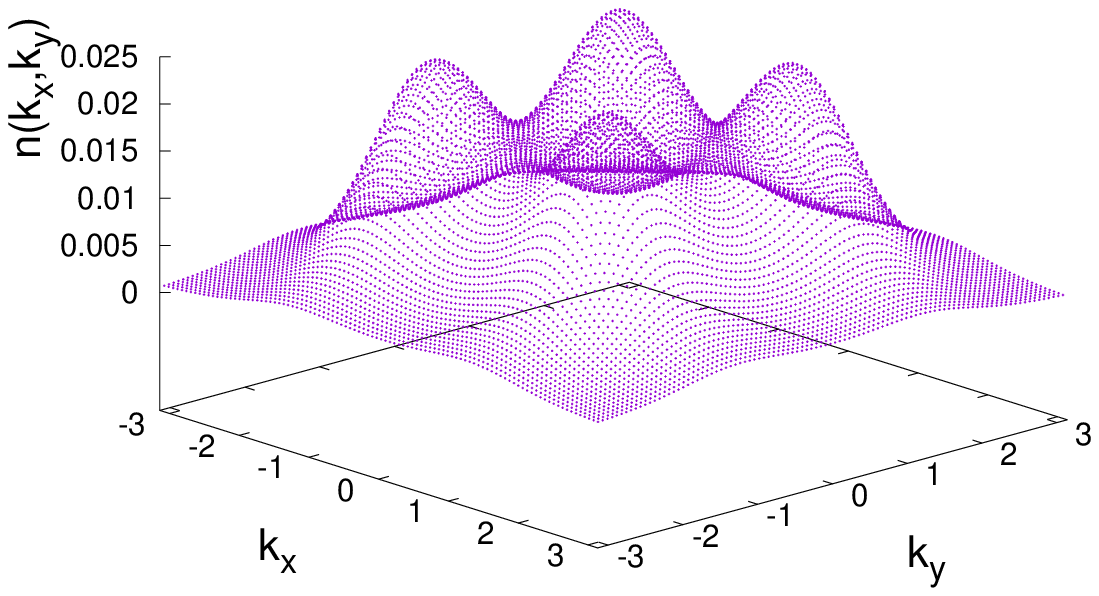}
           \label{fig4D}
        }
    \end{center}
    \caption{  (a) The non-interacting valence and conduction bands (relative to their corresponding chemical potentials) in the case of the
      band-structure given by Eq.~\ref{band.indirect} is shown for $\gamma=1.0$ eV  and bandwidth $W=$2 eV. (b), (c) and (d) illustrate the  calculated pairing gap, the quasiparticle energy (relative to the chemical potentials) and the ground-state momentum distribution of electrons in the conduction-band respectively, using $\epsilon=3$ and $a=5 a_0$. }
\vskip 0.2 in
\end{figure*}

\begin{figure*}
    \vskip 0.3 in \begin{center}
        \subfigure[]{
            \includegraphics[scale=0.6]{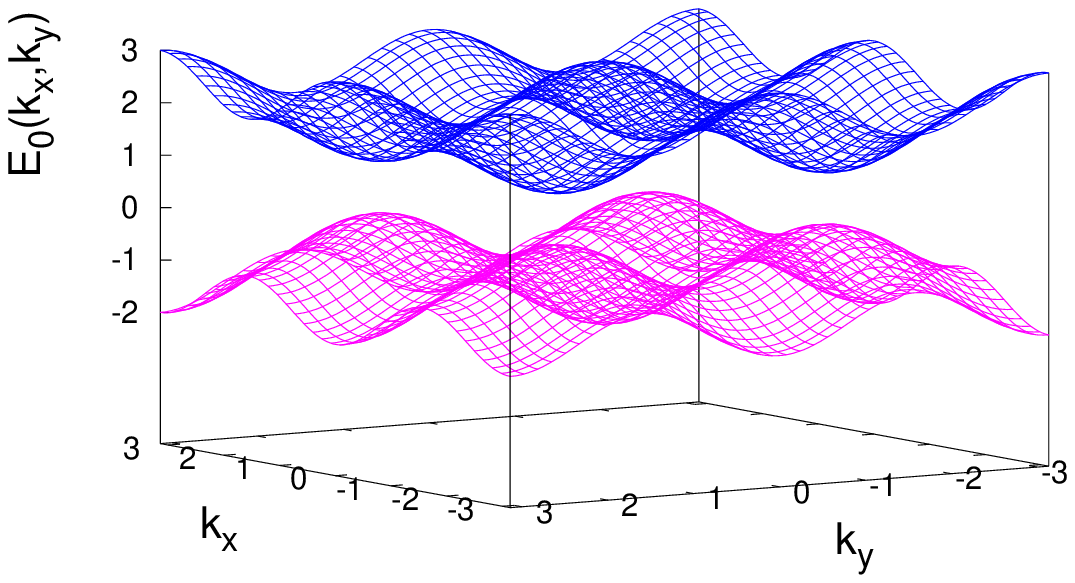}
           \label{fig5A}
        }
        \subfigure[]{
            \includegraphics[scale=0.6]{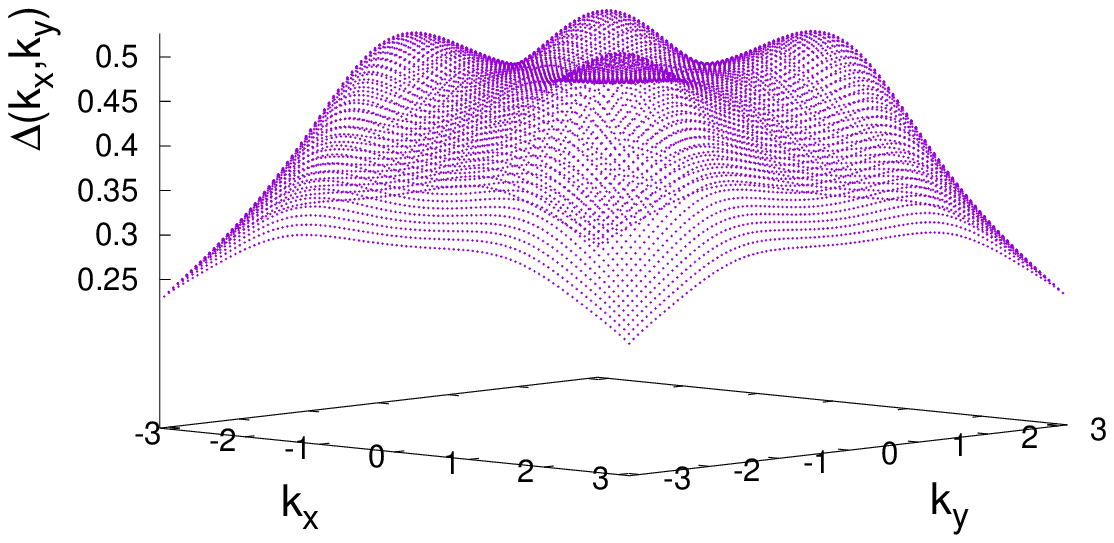}
            \label{fig5B}
         }
        \\
        \subfigure[]{
          \includegraphics[scale=0.6]{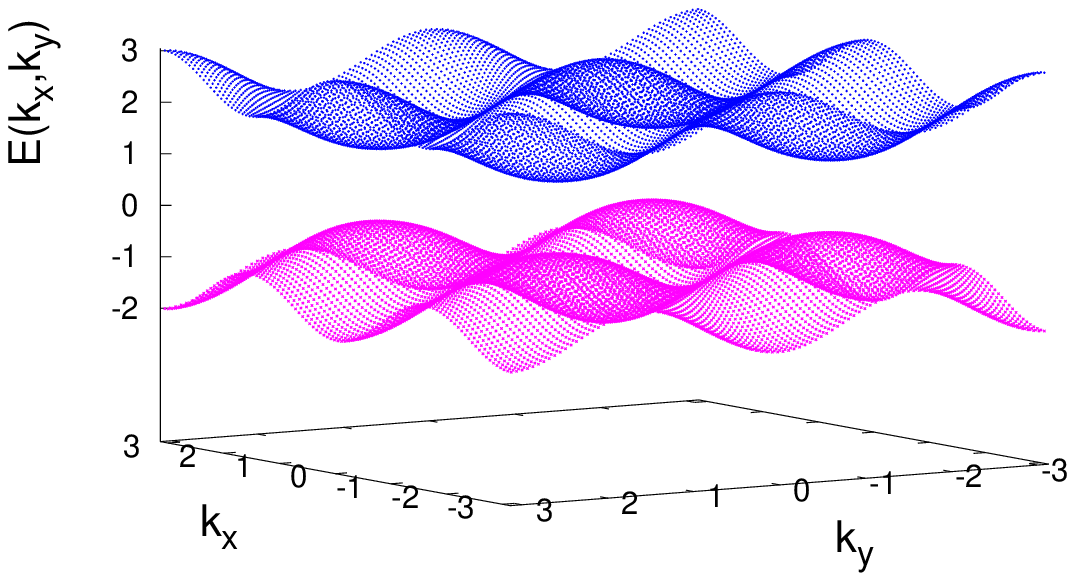}
           \label{fig5C}
        }
        \subfigure[]{
            \includegraphics[scale=0.6]{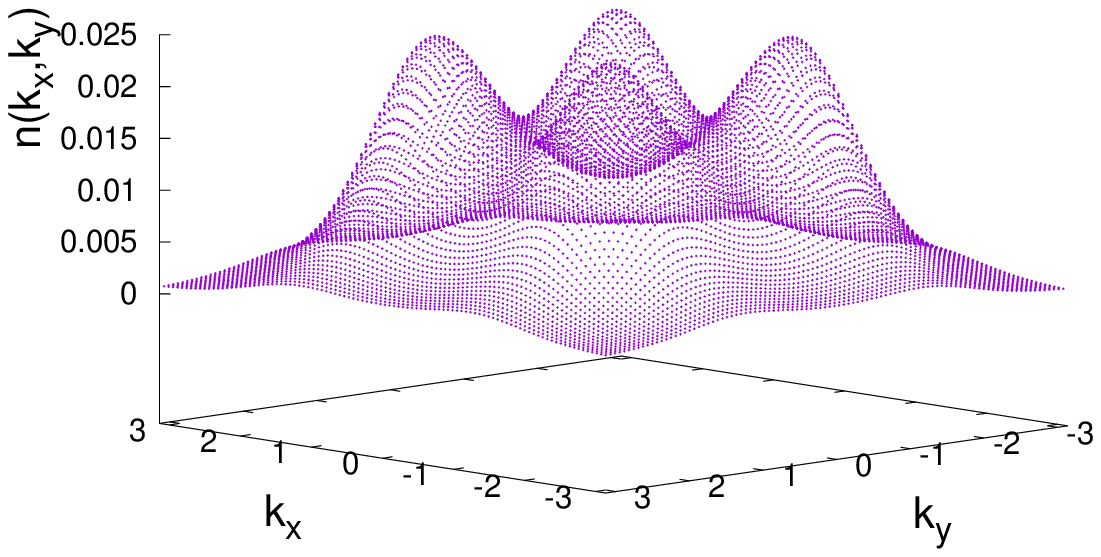}
           \label{fig5D}
        }
    \end{center}
    \caption{ (a) The non-interacting valence and conduction bands (relative to their corresponding chemical potentials) in the case of the
      band-structure given by Eq.~\ref{band-sym} is shown for $\gamma=1.0$ eV  and bandwidth $W=$2 eV. (b), (c) and (d) illustrate the  calculated pairing gap, the quasiparticle energy (relative to the chemical potentials) and the ground-state momentum distribution of electrons in the conduction-band respectively, using $\epsilon=3$ and $a=5 a_0$.}
\vskip 0.2 in
\end{figure*}

\section{General features and remarks}
\label{sec:features}

The approach discussed in the previous section is quite general and can be directly applied to real materials, using {\it ab initio} calculations. 
A band-structure calculation is required to determine the
valence and conduction band energies $E_{\alpha}(\vec k)$ and $E_{\beta}(\vec k)$
and their corresponding Bloch wavefunctions $|\psi_{\alpha\vec k}\rangle$ and
$|\psi_{\beta\vec k}\rangle$. The effective interaction $\hat V$
can be approximated using the random-phase-approximation (RPA) where $\hat V_{\alpha\beta} (\vec q)$
is the  Fourier transform of the Coulomb interaction screened by
the dielectric function.
The dielectric function in the RPA\cite{PhysRev.115.786,PhysRev.179.892} is given as
\begin{eqnarray}
  \epsilon(\vec q,\omega) &=& 1 - {{4 \pi e^2} \over {q^2}} \Pi_0(\vec q,\omega)\\
\Pi_0(\vec q,\omega)&\equiv&          {1 \over {\Omega}}
\sum_{\vec p \alpha\beta} 
   {{|\langle \psi_{\alpha\vec p}|e^{-i \vec q \cdot \vec r} | \psi_{\beta\vec p + \vec q} \rangle|^2} \over {E_{\beta}(\vec p + \vec q) - E_{\alpha}(\vec p) -\hbar \omega + i \eta}}, \nonumber \\
&\times& [F(E_{\beta}(\vec p + \vec q))-F(E_{\alpha}(\vec p))]    \label{RPA} 
  \end{eqnarray}
where $\Omega$ is the volume of the crystal and
$F(E)$ is the zero temperature Fermi-Dirac distribution and taking the
limit of $\eta \to 0$ correctly will yield the real part (i.e., the principal part of the
momentum space integral)  and the imaginary parts of $\epsilon(\vec q,\omega)$. 

Notice that as the conventional-insulator gap $G$ becomes smaller and smaller,
something that  has been widely discussed as a proper direction to take in order to enter the excitonic insulator phase, the dielectric function increases
and this weakens the attractive interaction between electron-hole pairs.
This works against the desired effect of producing a excitonic binding-energy greater than $G$. As we will see below the dependence of $\Delta(\vec k)$ on
$\epsilon$ close to the conventional to excitonic-insulator transition can be
stronger than its dependence on $G$. As we will show, this depends on
how far from the critical values of each of these parameters the material is.

One can naively try to make the same argument for the bandwidth, i.e.,
as the bandwidths of the conduction and valence bands decrease,
the denominator of Eq.~\ref{RPA} decreases and the
dielectric function should increase. However, this is not a
correct conclusion, because the matrix element  in the numerator 
also decreases, because the overlap between Wannier functions
would decrease as the bandwidth decreases.

In the present paper, we plan to restrict ourselves to a variety of
specific models which we would like to solve and which addresses
the following question, the answer of which should aid the experimental search
for an excitonic insulator: What are the relevant experimentally accessible
parameters  and how does the excitonic-insulator transition depend on each one of
them?
The electron-electron interaction in the insulator will be approximated by
the screened Coulomb interaction which involves only
a dielectric constant $\epsilon=\epsilon(\vec q=0,\omega=0)$, i.e.,
we ignore the momentum and frequency
dependence of the dielectric function and we only consider both the static and
the long-wavelength limit. Notice that $\epsilon$ is directly accessible to experiments and it is clear
that the candidate material to host the EI state should be selected to have an $\epsilon$ as small
as possible. For this reason, we will
restrict our models to two-dimensional materials where, because of the fact
that the material  is surrounded by vacuum, there is a good chance for the
material to have a small average value of $\epsilon$.
In addition, just because $\epsilon$ is accessible to experiments, it
will be an input parameter to our models and its microscopic determination will not be done here. As we will see below, there are a few other parameters
which are also relevant and accessible to experiment (and to electronic-structure calculations). 

\section{Solution of the gap equation for two-dimensional materials}
\label{sec:solution}

Next, we consider the  case of two-dimensional materials using the Rytova-Keldysh\cite{Rytova} potential for a slab:
\begin{eqnarray}
  V_{\mathrm{eff}}(q) ={1 \over A} {{2 \pi e^2} \over {\epsilon q} } {{e^{q L} + \delta}
    \over {e^{q L} -\delta}}, \hskip 0.2 in \delta = {{\epsilon -1} \over {\epsilon +1 }},
  \label{2d-potential}
\end{eqnarray}
where $A$ and $L$ are the area and the thickness of a slab, and $\epsilon$ is the dielectric constant.
We really do not know what the value of $L$ should be in a realistic case
of a system, and, in addition, we prefer not to have too many free parameters.
We will use a large enough value of $L=20 a_0$ ($a_0$ is the atomic unit of length, i.e., the Bohr radius)  where the results become
insensitive to values of $L$ greater than that.

In the next several subsections we will consider variations of
a two-band model, i.e., a single valence and single conduction band.
We will consider the cases of a direct and of an indirect effective gap in
Sec.~\ref{section:direct} and Sec.~\ref{section:indirect}.
We note that, while the real conventional-insulator gap $G$
is positive, the effective gap $\gamma \equiv G -\delta\mu$ can be tuned
to zero or to a negative value, when the chemical potential
for electrons and holes is effectively set to different values by the
laser illumination.
Therefore, in Sec.~\ref{section:negative}, we will examine the case where the effective gap $\gamma$ vanishes
and the case in which $\gamma$ becomes negative.
Lastly, in Sec.~\ref{section:multiple}, we will study the case where the
conduction band has
four minima at the finite momenta $\vec k = (\pm \pi/(2a),\pm \pi/(2a))$, while the valence band has maxima at these
points.

\subsection{Direct band-gap}
\label{section:direct}

First, we will consider the case of a direct gap using a single valence and and a single conduction band
parametrized by the following dispersion relations:
\begin{eqnarray}
  \epsilon_{\alpha}(\vec k) &=& 2 t_v (\cos(k_x a) + \cos(k_y a)-2),\nonumber \\
  \epsilon_{\beta}(\vec k) &=& \gamma -2 t_c (\cos(k_x a) + \cos(k_y a)-2).
  \label{simple.direct}
\end{eqnarray}
where $\gamma=G-\delta \mu$ is the effective insulating gap.
For simplicity and for avoiding using too many parameters, we will only
consider the case of $t_v=t_c$ and an $N\times N$ square lattice of area
$A = (N a)^2$ with periodic boundary conditions.
In Eq.~\ref{eq:gap}, the two-dimensional sum is over two independent integers
$m_x$ and $m_y$, which define the 2D vector $\vec k'$ from its components
$k'_x=m_x 2\pi/(Na)$ and $k'_y=m_y 2\pi/(Na)$, and each take
$N$ values, i.e., $-N/2 < m_{x,y} \le N/2$.

There is a range of the parameters, bandwidth $W = 8 t$, $\epsilon$,
$\gamma$ and $a$, where the gap equation, i.e., Eq.~\ref{eq:gap},
has non-trivial solutions. This range and the transition to
the excitonic insulator phase is investigated in Sec.~\ref{EItransition}.
In Fig.~\ref{fig1A}, the non-interacting band-structure is shown for
the case of a effective gap $\gamma=1.0$ eV and $t=0.25$ eV.
For this case,  the self-consistent solution to the gap equation, i.e., to Eq.~\ref{eq:gap},
is shown in  Fig.~\ref{fig1B} using for dielectric constant $\epsilon=3$ and
$a=5 a_0$ ($a_0$ is the Bohr radius).
Notice that the gap has significant momentum dependence and
is larger at the $\Gamma$ point where the conduction (valence) band
has its minimum (maximum).
In Fig.~\ref{fig1C} we present 
the two quasiparticle bands resulting for this case. Notice that
they have a similar shape to the non-interacting conventional-insulator bands shown in Fig.~\ref{fig1A}. The most significant difference
is in the vicinity of the $\Gamma$ point where the gap opens wider
because the gap-function $\Delta(k_x,k_y)$ is larger near the $\Gamma$ point.
In Fig.~\ref{fig1D} we present the ground-state electron momentum  distribution
(i.e., $n_{\beta}(\vec k)$ given by Eq.~\ref{eq:md}). The hole momentum distribution  is the same for this case because the valence and conduction bands are mirror symmetric with respect to
the Fermi energy. 
Notice that its peak is at the $\Gamma$ point as expected.

\subsection{Zero and negative effective non-interacting gap}
\label{section:negative}

We used the band-structure given by Eq.~\ref{simple.direct} to investigate
the case where the effective gap  $\gamma$ is zero or negative. 
In the case where the effective insulating gap $\gamma$ of the non-interacting band-structure
is zero and the chemical potentials for both bands are set to zero, we expect
to find particle-hole pairing condensate for all values of the
dielectric constant and of the rest of the other parameters.  This is the well-known instability of the Fermi-liquid system when there is an
attractive interaction present between  fermionic species.
As the value of $\epsilon$ becomes larger and larger the value of the gap will become smaller and smaller but never zero.

In Fig.~\ref{fig2A}, the non-interacting band-structure is shown for
$t=0.25$ eV and $\gamma=0$.
For this case,  the self-consistent solution to the gap equation, i.e., to Eq.~\ref{eq:gap},
is shown in  Fig.~\ref{fig2B} for $\epsilon=10$ and $a =20 a_0$.
In Fig.~\ref{fig2C} 
the two quasiparticle bands are shown for the case of these parameters values. Notice that the gap which opens near the $\Gamma$ point is too small,
i.e., with a maximum value $\Delta(0,0) \sim 0.008$ eV, to be
visible on the scale used in Fig.~\ref{fig2C}.
In Fig.~\ref{fig2D} we present the ground-state distribution
of electrons in the conduction-band which has a sharp peak at the $\Gamma$ point.

In Fig.~\ref{fig3A}, the non-interacting band-structure is shown for
the case of a negative effective gap $\gamma=-0.5$ eV and $t=0.25$ eV.
For this case,  the self-consistent solution to the gap equation, i.e., to Eq.~\ref{eq:gap},
is shown in  Fig.~\ref{fig3B} for $\epsilon=10$ and $a=20 a_0$.
Notice the small-size of the EI gap and this is so because the
dielectric constant is large. In addition, notice the singularities along
the nodal line of the Fermi surface caused
by the crossing of the two non-interacting bands.
Fig.~\ref{fig3C} illustrates
the two quasiparticle bands           for this case.
In Fig.~\ref{fig3D} we present the ground-state electron momentum distribution,
which has a drumhead shape with its circular edge defined by the Fermi surface nodal line.

\subsection{Indirect band-gap}
\label{section:indirect}
To investigate the presence of a non-trivial solution to the
gap equation when the band-structure has an indirect gap, we consider the
following simple band-structure:
\begin{eqnarray}
  \epsilon_{\alpha}(\vec k) &=& -2 t_v (\cos(2 k_x a) + \cos(2 k_y a)+2),\nonumber \\
  \epsilon_{\beta}(\vec k) &=& \gamma -2 t_c (\cos(k_x a) + \cos(k_y a)-2).
  \label{band.indirect}
\end{eqnarray}
These bands, are shown in Fig.~\ref{fig4A}
using an effective gap $\gamma=1.0$ eV and $t_c=t_v=0.25$ eV.
Notice that the conduction band has a minimum at the $\Gamma$ point, while the
valence band has a minimum at the $\Gamma$ point and four maxima at $(\pm {{\pi} \over {2 a}}, \pm {{\pi}\over {2 a}})$. Therefore, the indirect gap is from
any of these four point of the valence band to the conduction-band energy at the
$\Gamma$ point.
For this case,  the self-consistent solution to the gap equation
is shown in  Fig.~\ref{fig4B} for $\epsilon=3$ and $a=5 a_0$.
Notice that $\Delta(\vec k)$ is significantly reduced compared to the
case of the direct gap. This is expected because in the indirect gap
case the particle-hole density of states for a particle at $\vec k$ and a hole at $-\vec k$ is significantly reduced in this case for the
lowest energy excitations. The two quasiparticle bands are shown in Fig.~\ref{fig4C}  for this case.
In Fig.~\ref{fig4D} the ground-state momentum
distribution of electrons in conduction-band is presented.
Remarkably, there are four peaks at the valence-band maxima; the reason for
this is that the energy-gap for a zero-momentum transfer excitation is minimum
at the four $\vec k$ points where the peaks are formed.
Because the 
density of states of the conduction-band is finite
at those same four $\vec k$ points, the particle-hole density of states for a particle at $\vec k$ and a hole at $-\vec k$  is reduced as
compared to the direct-gap case. In the latter case, the density of
states diverges at the $\Gamma$ point in both the conduction and valence bands.

\subsection{Multiple direct band-gaps}
\label{section:multiple}

Here, we examine the case of a band-structure with multiple conduction-band minima
which coincide with the valence-band maxima.  Shown in 
 Fig.~\ref{fig5A} is an example of such a scenario, where
the non-interacting band-structure is given by:
\begin{eqnarray}
  \epsilon_{\alpha}(\vec k) &=& 4 t_v (\sin^2(k_x a) + \sin^2(k_y a)+2),\nonumber \\
  \epsilon_{\beta}(\vec k) &=& \gamma -4 t_c (\sin^2(k_x a) + \sin^2(k_y a)-2).
  \label{band-sym}
\end{eqnarray}
We have used an effective gap $\gamma=1.0$ eV and $t_c=t_v=0.25$ eV.
There are four such extrema of the conduction and valence band
at $(\pm \pi/(2a), \pm\pi/(2a))$.
For this case,  the self-consistent solution to the gap equation
is shown in  Fig.~\ref{fig5B} for $\epsilon=3$ and $a=5 a_0$ where $\Delta(k_x,k_y)$ exhibits four maxima at those band extrema.
In Fig.~\ref{fig5C} the two quasiparticle bands are
illustrated. Notice that, while the momentum dependence is similar to that of
the non-interacting bands,
the gap in the bands in the EI state is  larger than
that of the non-interacting
conventional-insulator case.
The ground-state momentum distribution of electrons
occupying the conduction-band is presented in Fig.~\ref{fig5D} and exhibits
four rather well-defined peaks at the same momenta.
The density of  states of the conduction band diverge
at the same four $\vec k$ points as the valence band, therefore, the particle-hole density of states for a particle at $\vec k$ and a hole at $-\vec k$ 
is very large at each one of those four points.

\begin{figure*}
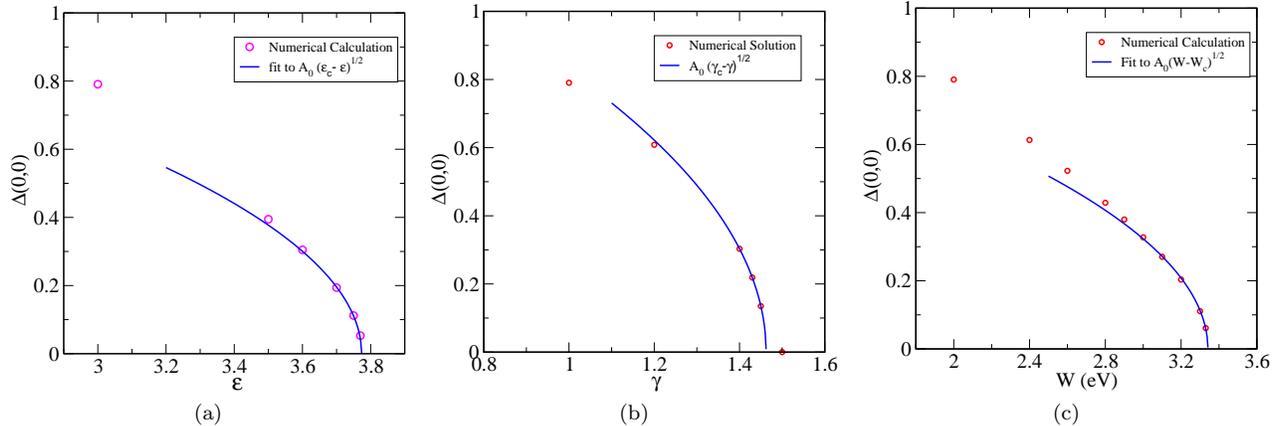

    \vskip 0.3 in \begin{center}
        \subfigure[]{
            \includegraphics[scale=0.3]{Fig6a}
           \label{fig6a}
        }
        \subfigure[]{
            \includegraphics[scale=0.3]{Fig6b}
            \label{fig6b}
        } 
        \subfigure[]{
          \includegraphics[scale=0.3]{Fig6c}
          \label{fig6c}
        }
    \end{center}
    \caption{ (a) The dependence of the excitonic gap $\Delta(0,0)$
              on $\epsilon$ using for bandwidth $W$, $\gamma$, and $a$, 2 eV, 1 eV and $5 a_0$
              respectively. The solid blue-line gives the result of the fit to the
              form of Eq.~\ref{epsilon.fit} which yields $A_0=0.721275$ and $\epsilon_c=3.77511$. (b) The dependence of  $\Delta(0,0)$
              on $G$ by fixing the
              parameters, bandwidth $W$, $\epsilon$, and $a$ to 2 eV, 3 and  $5 a_0$
            respectively. The solid blue-line gives the result of the fit to the
            form of Eq.~\ref{gamma.fit} which yields $B_0=1.21435$ and $\gamma_c=1.46238$. (c) The dependence of  $\Delta(0,0)$
              on the bandwidth $W$ by fixing the
              parameters, bandwidth $G$, $\epsilon$, and $a$ to 1 eV, 3 and $5 a_0$
            respectively. The solid blue-line gives the result of the fit to the
            form of Eq.~\ref{W.fit} which yields $C_0=0.552829$ and $W_c=3.34156$.}
              \vskip 0.2 in
\end{figure*}

\section{Conventional to excitonic insulator transition}
\label{EItransition}

Here, we discuss the results of our investigation of the presence of a non-trivial solution $\Delta (\vec k)$
to the gap equation, i.e., to Eq.~\ref{eq:gap}, for the case of direct and positive effective
gap $\gamma$ using the simple
dispersion given by the Eq.~\ref{simple.direct}. We investigate the dependence
of the EI gap  on the parameters $\epsilon$, $\gamma$ and the bandwidth $W$.

Figs.~\ref{fig6a},\ref{fig6b},\ref{fig6c} illustrate the dependence of $\Delta (\vec k=0)$ on each one of the parameters $\epsilon$, $\gamma$ and $W$
by fixing the values of other two.
The solid blue lines represent  fits to the formulae
\begin{eqnarray}
  \Delta(0,0) &=& A_0 (\epsilon_c - \epsilon)^{1/2},
  \label{epsilon.fit}\\
    \Delta(0,0) &=& B_0 (\gamma_c - \gamma)^{1/2},
  \label{gamma.fit}\\
  \Delta(0,0) &=& C_0 (W_c - W)^{1/2},
  \label{W.fit}
\end{eqnarray}
near the critical values $\epsilon_c$, $\gamma_c$ and $W_c$,  as expected because of the mean-field character of the solution.
The coefficients and the critical values are given in the caption of that figure.
Notice that the agreement with the mean-field order-parameter critical exponent 1/2 is quite good near the critical point.

We note that for $\gamma=1.0$ eV the required value of $\epsilon$
is less than 3.8. However, as the value of $\gamma$ or $W$ decreases
the value of $\epsilon_c$ becomes larger. For example, if we use
$\gamma = 0.5$ eV and keep the rest of the parameters ($W=2$ eV and $a = 5 a_0$)
the same, we obtain $\epsilon_c \sim 5.6$.
We note that the value of the self-consistent solution for the gap
does not depend independently on the parameters $\epsilon$ and $a$.
It depends on the product $\epsilon a$. So, the combination of the unit-cell size and $\epsilon$ has to be
taken into consideration when searching for a good candidate material
to realize the EI state.

\section{Discussion and conclusions}
\label{sec:Discussion}

\begin{figure}
            \includegraphics[scale=0.3]{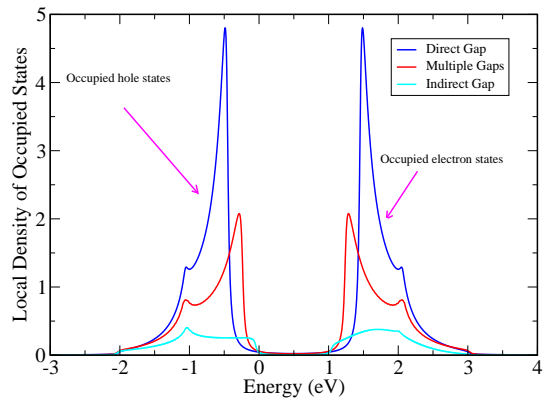}
            \label{LDOS}
            \caption{The zero-temperature local density of occupied states for the various cases of
              band-structure discussed here. Tunneling microscopy can
              measure the LDOS which can provide a smoking gun for the
            presence of the EI state.}
  \end{figure}

We have studied the transition from the conventional-insulator state
to that of an excitonic insulator in
several cases of 2D band-structures with various different features.
We find that the band-structures with a direct-gap (keeping all other
band characteristics and $\epsilon$ the same) yield a larger EI gap as compared to
the case of an indirect gap. The EI state in the
case of an indirect gap or in the case of
band-structures with direct gaps at wavevectors different than
the $\Gamma$-point, has a zero-temperature electron momentum distribution
with peaks at such non-zero wavevectors. 

The electron momentum distribution
is experimentally accessible by tunneling experiments. Tunneling experiments
measure the local density of occupied states (LDOS) which, in our case, is
given as
\begin{eqnarray}
  \rho(\omega) = {1 \over {N}}\sum_{\vec k,\nu=\alpha,\beta} n_{\nu}(\vec k)
  \delta(\hbar \omega - {\cal E}_{\nu}).
\end{eqnarray}
Therefore, $\rho(\omega)$ can be calculated using the quasiparticle bands and the momentum
distributions given by Eq.~\ref{eq:md} and is illustrated in Fig.~\ref{LDOS}
for the cases of direct gap, indirect gap
and multiple gaps discussed in Sec.~\ref{section:direct}, Sec.~\ref{section:indirect} and Sec.~\ref{section:multiple} respectively. In addition, if one is able to
accurately measure the 
differential tunnel conductance as a function of location
within the unit-cell of the Bravais lattice, then
the electron/hole ground-state
momentum distribution can be inferred\cite{doi:10.1073/pnas.2207449119,PhysRevB.101.085142}.

Therefore, the zero-temperature electron momentum distribution, through
the corresponding observable LDOS,
can be used
as a criterion for the presence of the EI state. In the 
case of insulators with an indirect gap or in materials with
multiple direct-gap band-structures the momentum distribution
develops peaks at  non-zero $\vec k$-points in the
Brillouin zone upon entering the EI state. These features can also
be seen experimentally, if the LDOS is probed by measuring
the differential tunnel conductance as a function of location and
tip-sample voltage\cite{doi:10.1073/pnas.2207449119,PhysRevB.101.085142}.

In the case of a direct-gap band-structure, we also studied the dependence
of the non-trivial gap function solution, $\Delta(k_x,k_y)$,  on the various parameters, such as, the band-structure bandwidth and effective band-gap, as well its dependence
on the dielectric constant of the 2D material. We find that the dependence
of $\Delta(0,0)$ on parameters, such as, the dielectric constant
$\epsilon$, the bandwidth $W$, and the effective
band-gap $\gamma$ near their corresponding critical values, is consistent
with mean-field critical behavior as expected. More importantly, we
find that conventional 2D insulators with not too-small gap (i.e., $\gamma \sim 1$ eV) and not too-small bandwidth (i.e., $W \sim 2$ eV)
can host an EI state, provided that the dielectric constant is
smaller than approximately $4$. Materials with larger values of $\epsilon$ must
have smaller effective $\gamma$ and/or $W$.

Our results
and conclusions should be relevant to real quasi-2D materials and
superlattices.
TMD monolayers or multilayers are structurally stable and display a variety of
band gaps and dielectric properties. Those with smaller dielectric constant,
smaller gap and narrow
conduction and valence bands should be more promising to
realize the excitonic insulator.
Furthermore, the organic–inorganic hybrid perovskites\cite{Stoumpos2016,doi:10.1126/science.aac7660,Gao2019} might be a suitable class of materials to
search for a potential realization of the EI state. These
insulators form a superlattice
structure where the perovskite layers are separated from each other
by means of organic molecules. One should select the organic molecules
to be among those with relatively small dielectric constant and
the atoms which form the halide-perovskite layer should be selected
to yield a relatively small gap with as narrow as possible valence and conduction bands nearest to the Fermi level.

\section{Acknowledgment}
I would like to thank Hanwei Gao for useful interactions.
This work was supported 
by the U.S. National Science
Foundation under Grant No. NSF-EPM-2110814.

\newpage
\appendix
\section{Diagonalization of the Bogoliubov-de Gennes matrix}
\label{AppendixA}
Next we diagonalize the matrix ${\bf M}_{\alpha\beta}(\vec k)$ to find
its eigenstates and its corresponding eigenvalues.
We notice that the matrix ${\bf M}$ has the following form:
\begin{eqnarray}
 {\bf M} =  \begin{pmatrix}
    {\bf D}_0 & {\bf T}^* \\
    -{\bf T} & - {\bf D}_0
\end{pmatrix},
\end{eqnarray}
where ${\bf T}$ and ${\bf D}_0$ are the following $2 \times 2$ matrices:
\begin{eqnarray}
 {\bf D}_0 =  \begin{pmatrix}
    -\epsilon_{\alpha}(\vec k) & 0 \\
    0 &  \epsilon_{\beta}(\vec k)
\end{pmatrix}, \hskip 0.01 in 
 {\bf T} =  \begin{pmatrix}
    0 & -\Delta_{\alpha\beta}(\vec k) \\
    \Delta_{\alpha\beta}(\vec k) &  0
\end{pmatrix}
\end{eqnarray}
and ${\bf T}^*$ is the complex conjugate (not the adjoint) of the matrix ${\bf T}$. The diagonalization equation:
\begin{eqnarray}
  \begin{pmatrix}
    {\bf D}_0 & {\bf T}^* \\
    -{\bf T} & - {\bf D}_0
  \end{pmatrix} \begin{pmatrix} {\bf u} \\ {\bf v} \end{pmatrix} = {\cal E}_{\lambda}
  \begin{pmatrix} {\bf u} \\ {\bf v} \end{pmatrix},
\end{eqnarray}
splits into the following two matrix equations:
\begin{eqnarray}
  {\bf D}_0 {\bf u} + {\bf T}^* {\bf v} &=& {\cal E}_{\lambda} {\bf u},\label{eq1} \hskip 0.2 in
  {\bf D}_0 {\bf v} + {\bf T} {\bf u} = - {\cal E}_{\lambda} {\bf v}. \label{eq2}
\end{eqnarray}
Now, Eq.~\ref{eq1} yields
\begin{eqnarray}
  {\bf v} = ({\bf T^*})^{-1} ({\cal E}_{\lambda} -{\bf D}_0) {\bf u}.
  \label{eq-v}
\end{eqnarray}
The inverse of the matrix ${\bf T}^*$ is
\begin{eqnarray}
  ({\bf T}^*)^{-1} = {1 \over {\Delta^*_{\alpha\beta}(\vec k)}} 
  \begin{pmatrix} 0 & 1\\
    -1  & 0 \end{pmatrix},
\end{eqnarray}
Now, using Eq.~\ref{eq-v} and by writing
\begin{eqnarray}
  {\bf u} = \begin{pmatrix} u_1 \\ u_2 \end{pmatrix}, \hskip 0.2 in 
  {\bf v} = \begin{pmatrix} v_1 \\ v_2 \end{pmatrix},
    \end{eqnarray}
we find that
\begin{eqnarray}
  v_1 = -{{\epsilon_{\beta}(\vec k) - {\cal E}_{\lambda}} \over {\Delta^*_{\alpha\beta}(\vec k)}} u_2, \hskip 0.2 in 
  v_2 = -{{\epsilon_{\alpha}(\vec k) + {\cal E}_{\lambda}} \over {\Delta^*_{\alpha\beta}(\vec k)}} u_1,  \label{eq:v1v2}
\end{eqnarray}
Now, Eq.~\ref{eq2} yields
\begin{eqnarray}
  {\bf u} = -({\bf T})^{-1} ({\cal E}_{\lambda} +{\bf D}_0) {\bf v}
  \label{eq-u}
\end{eqnarray}
The inverse of the matrix ${\bf T}^*$ is
\begin{eqnarray}
  ({\bf T})^{-1} = {1 \over {\Delta_{\alpha\beta}(\vec k)}} 
  \begin{pmatrix} 0 & 1\\
    -1  & 0 \end{pmatrix}, 
\end{eqnarray}
which lead to the following relations:
\begin{eqnarray}
  u_1 = -{{\epsilon_{\beta}(\vec k) + {\cal E}_{\lambda}} \over {\Delta_{\alpha\beta}(\vec k)}} v_2, \hskip 0.2 in 
  u_2 = -{{\epsilon_{\alpha}(\vec k) - {\cal E}_{\lambda}} \over {\Delta_{\alpha\beta}(\vec k)}} v_1, \label{eq:u1u2}
\end{eqnarray}

Eqs~\ref{eq:u1u2},\ref{eq:v1v2} have to be simultaneously satisfied. This leads to the
following equations:
\begin{eqnarray}
v_1 &=& {{\epsilon_{\alpha}-{\cal E}_{\lambda}} \over {\Delta_{\alpha\beta}(\vec k)}}
  {{\epsilon_{\beta}-{\cal E}_{\lambda}} \over {\Delta^*_{\alpha\beta}(\vec k)}} v_1, \\
v_2 &=& {{\epsilon_{\alpha}+{\cal E}_{\lambda}} \over {\Delta_{\alpha\beta}(\vec k)}}
{{\epsilon_{\beta}+{\cal E}_{\lambda}} \over {\Delta^*_{\alpha\beta}(\vec k)}} v_2.
\end{eqnarray}
However, if both $v_1$ and $v_2$ are assumed to be non-zero, then, this leads
to 
\begin{eqnarray}
 (\epsilon_{\alpha}-{\cal E}_{\lambda})
 (\epsilon_{\beta}-{\cal E}_{\lambda}) =
 (\epsilon_{\alpha}+{\cal E}_{\lambda})
 (\epsilon_{\beta}+{\cal E}_{\lambda}),
\end{eqnarray}
i.e.,  to ${\cal E}_{\lambda} = 0$, which can only be satisfied by the
trivial solution $v_1=v_2=u_1=u_2$, because of the structure of the matrix
${\bf M}$.

However, we have the alternative solutions that either a) $v_1\ne0$ and $v_2= 0$
or b) $v_2\ne 0$ and $v_1= 0$.

a) $  v_2 = 0 \to u_1 = 0$, and using Eqs~\ref{eq:u1u2} we obtain
\begin{eqnarray}
  {{(\epsilon_{\beta}(\vec k) - {\cal E}_{\lambda})
    (\epsilon_{\alpha}(\vec k) - {\cal E}_{\lambda})}
    \over {|\Delta_{\alpha\beta}(\vec k)|^2}} = 1.
  \end{eqnarray}
This equation determines the eigenvalues in this case to be given as
\begin{eqnarray}
  {\cal E}_{1} &=& \bar{\epsilon}_{\alpha\beta} 
  + {\cal R}_{\alpha\beta},
  \hskip 0.2 in {\cal E}_{2} = \bar{\epsilon}_{\alpha\beta}  - {\cal R}_{\alpha\beta}, \\
   {\cal R}_{\alpha\beta} &\equiv& 
  \sqrt{( \delta \epsilon_{\alpha\beta} )^2 + |\Delta_{\alpha\beta}(\vec k)|^2}, \\
  \bar{\epsilon}_{\alpha\beta} &\equiv& {{\epsilon_{\alpha}(\vec k) + \epsilon_{\beta}(\vec k)} \over 2}, \hskip 0.1 in
  \delta \epsilon_{\alpha\beta} \equiv {{\epsilon_{\alpha}(\vec k) - \epsilon_{\beta}(\vec k)} \over 2}. 
  \end{eqnarray}
In this case the non-zero coefficients are related as follows
\begin{eqnarray}
  u_2 = -{{\epsilon_{\alpha}-{\cal E}_{\lambda}} \over {\Delta_{\alpha\beta}(\vec k)}} v_1
  \end{eqnarray}
and after normalization (and choosing an overall phase factor) the corresponding eigenvectors are obtained as
\begin{eqnarray}
  b^{(1,2)}_{\vec k} &=& u^{(1,2)}_{2} c^{\dagger}_{\vec k \beta} + v^{(1,2)}_{1} h_{-\vec k\alpha}, \label{eq:quasi1} \\
  u^{(1)}_{2} &=& -{{\delta\epsilon_{\alpha\beta} -{\cal R}_{\alpha\beta}}
    \over {\Delta_{\alpha\beta}}} v_1^{(1)}, \\
    u^{(2)}_{2} &=&-{{\delta\epsilon_{\alpha\beta} +{\cal R}_{\alpha\beta}}
      \over {\Delta_{\alpha\beta}}} v_1^{(2)}.
\end{eqnarray}

b) $  v_1 = 0 \to u_2 = 0$. In this case using Eqs~\ref{eq:v1v2} 
\begin{eqnarray}
  {{(\epsilon_{\beta}(\vec k) + {\cal E}_{\lambda})
    (\epsilon_{\alpha}(\vec k) + {\cal E}_{\lambda})}
    \over {|\Delta_{\alpha\beta}(\vec k)|^2}} = 1.
  \end{eqnarray}
The latter equation determines the eigenvalues in this case to be given as
\begin{eqnarray}
  {\cal E}_3 = - {\cal E}_1, \hskip 0.3 in 
  {\cal E}_4 = - {\cal E}_2. \end{eqnarray}
After normalization (and choosing an overall phase factor) the corresponding eigenvectors are obtained as
\begin{eqnarray}
  b^{(3,4)}_{\vec k } &=& u^{(3,4)}_{1} h^{\dagger}_{-\vec k\alpha}, + v^{(3,4)}_{2} c_{\vec k \beta}, \label{eq:quasi2}\\
  v^{(3)}_{2 } &=& -{{\delta\epsilon_{\alpha\beta} -{\cal R}_{\alpha\beta}}
    \over {\Delta^*_{\alpha\beta}}} u_1^{(3)}, \\
    v^{(4)}_{2} &=& -{{\delta\epsilon_{\alpha\beta} +{\cal R}_{\alpha\beta}}
      \over {\Delta^*_{\alpha\beta}}} u_1^{(4)}.
\end{eqnarray}
In summary, after normalizing the coefficients, we obtain the following
solutions
\begin{eqnarray}
  {\cal C}^{\dagger}_{\vec k} &=& \kappa_{-} c^{\dagger}_{\vec k \beta} - \chi_- h_{-\vec k\alpha}, \label{eq:quasia} \\
  {\cal H}_{-\vec k} &=& \kappa_{+} c^{\dagger}_{\vec k \beta} - \chi_+  h_{-\vec k\alpha}, \label{eq:quasia} \\
  \kappa_{\pm} &=& {{\delta\epsilon_{\alpha\beta} \pm {\cal R}_{\alpha\beta}}
    \over {{\cal D}^{\pm}_{\alpha\beta}}}, \hskip 0.2 in 
    \chi_{\pm} = {{\Delta_{\alpha\beta}}     \over {{\cal D}^{\pm}_{\alpha\beta}}}.\\
       {\cal D}^{\pm}_{\alpha\beta} &\equiv& \sqrt{(\delta\epsilon_{\alpha\beta}\pm{\cal R}_{\alpha\beta})^2+|\Delta_{\alpha\beta}|^2},
\end{eqnarray}
where ${\cal C}^{\dagger}_{\vec k}$ corresponds to the ${\cal E}_1$ eigenvalue and
  when $\Delta \to 0$, the operator ${\cal C}^{\dagger}_{\vec k} \to c^{\dagger}_{\vec k \beta}$.
  The ${\cal H}_{-\vec k}$ operator corresponds to the ${\cal E}_2$
    eigenvalue and
  when $\Delta \to 0$, ${\cal H}_{-\vec k} \to h_{-\vec k \beta}$.
  The other two solutions which correspond to the $-{\cal E}_1$ and $-{\cal E}_2$ eigenvalues are the ${\cal C}_{\vec k}$ and ${\cal H}^{\dagger}_{-\vec k}$ operators
  which are the adjoint operators. 
  
We need to reverse the
Eq.~\ref{eq:quasi1} and Eq.~\ref{eq:quasi2} to determine
the original electron/hole operators in terms of the quasiparticle
operators. We find that:
\begin{eqnarray}
  c^{\dagger}_{\vec k\beta} &=& {{{\cal D}^+_{\alpha\beta}} \over {2 {\cal R}_{\alpha\beta}}} {\cal H}_{-\vec k} - { {{\cal D}^{-}_{\alpha\beta}} \over {2 {\cal R}_{\alpha\beta}}}   {\cal C}^{\dagger}_{\vec k}, \\
  h_{-\vec k \alpha} &=& -\xi^{+} {\cal C}^{\dagger}_{\vec k} + \xi^- {\cal H}_{-\vec k},\\
\xi^{\pm} &=& {{(\delta \epsilon_{\alpha\beta} \pm {\cal R}_{\alpha\beta}) {\cal D}^{\mp}_{\alpha\beta}} \over {2 {\cal R}_{\alpha\beta} \Delta_{\alpha\beta}}}.
\end{eqnarray}

\end{document}